\documentclass[conference,twocolumn]{IEEEtran}

\ifCLASSINFOpdf
\else
\fi

\usepackage{setspace}
\usepackage{graphicx}
\usepackage{amsthm}
\usepackage{subfigure}

\newtheorem{Lemma}{Lemma}

\begin{document}

\title{A Novel Relay-Aided Transmission Scheme in Cognitive Radio Networks}

\author
{\IEEEauthorblockN{Wael Jaafar}
\IEEEauthorblockA{\'{E}cole Polytechnique de Montr\'{e}al\\Department of Electrical Engineering\\
Montreal, Canada\\
Email: wael.jaafar@polymtl.ca}
\and
\IEEEauthorblockN{Wessam Ajib}
\IEEEauthorblockA{Universit\'{e} du Qu\'{e}bec \`{a} Montr\'{e}al\\
Department of Computer Science\\Montreal, Canada\\
Email: ajib.wessam@uqam.ca}
\and
\IEEEauthorblockN{David Haccoun}
\IEEEauthorblockA{\'{E}cole Polytechnique de Montr\'{e}al\\
Department of Electrical Engineering\\
Montreal, Canada\\
Email: david.haccoun@polymtl.ca} }

\maketitle


\begin{abstract}
In underlay cognitive radio networks, unlicensed secondary users are allowed to share the spectrum with licensed primary users when the interference induced on the primary transmission is limited. In this paper, we propose a new cooperative transmission scheme for cognitive radio networks where a relay node is able to help both the primary and secondary transmissions. We derive exact closed-form and upper bound expressions of the conditional primary and secondary outage probabilities over Rayleigh fading channels. Furthermore, we proposed a simple power allocation algorithm. Finally, using numerical evaluation and simulation results we show the potential of our cooperative transmission scheme in improving the secondary outage probability without harming the primary one.
 \end{abstract}

\IEEEpeerreviewmaketitle

\section{Introduction}
Cognitive Radio (CR) is an interesting technology that allows the access for unlicensed secondary users to communicate in parts of the licensed spectrum bands when they are not in use by the licensed primary users \cite{Mitola}-\cite{Haykin}.
Cooperative diversity, proposed and studied in \cite{Laneman1}\nocite{Erkip_part12}-\cite{Jaafar4}, has the advantage to improve the spatial diversity order when one or many relay nodes participate in the communication.

Integrating cooperative diversity within CR led to new research interests such as cooperative spectrum sensing \cite{Letaief} and cooperative transmission \cite{Simeone1}\nocite{Simeone2,Han}-\cite{Zou} in Cognitive Relay Networks (CRNs). In \cite{Letaief}, the authors showed the importance of cooperative diversity to circumvent with the hidden terminal problem. The authors in \cite{Simeone1}-\cite{Simeone2} assumed that a secondary user can act as a relay node for primary transmissions. The main benefits are reducing the delay of the primary transmissions and increasing the access opportunities to the licensed spectrum band for the secondary users. In \cite{Han}, the authors proposed a cooperative scheme where the secondary transmitter sends the primary signal along with its own signal without affecting the primary outage probability. The authors derived a critical distance between the primary and the secondary transmitters for which some given fraction of the transmit power is chosen at the secondary transmitter to forward the primary signal. As a consequence, the primary outage probability is respected and a secondary access is achieved. Cooperation among secondary users has also been studied. In \cite{Zou}, the authors proposed a cooperative scheme where a secondary user is optimally selected to act as a relay for an ongoing secondary transmission leading to the secondary outage probability significantly improved compared to non-cooperative access.

\begin{figure}
  \centering
\includegraphics[width=210pt]{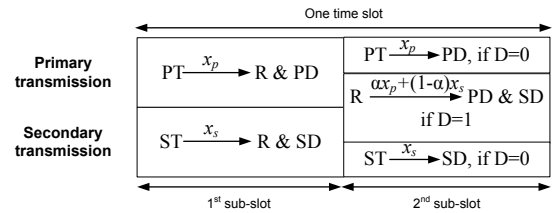}
  \caption{Illustration of the transmission using the proposed scheme in cognitive radio networks}
  \label{Fig2_Network_R_assisted_PU}
\end{figure}

Most of the previous research activities that investigated cooperative diversity in CRNs has considered assisting the primary transmission or the secondary transmission and none of them considered assisting both the primary and secondary transmissions simultaneously. In this paper, we propose a novel cooperative scheme whereby a non-cognitive relay node helps simultaneously the primary and the secondary transmissions, taking into account the existing interfering links between the primary and the secondary transmissions. Since primary and secondary transmissions will benefit from the proposed cooperative transmission scheme, the relay node may be either a primary or a secondary user.

The main advantages of the proposed cooperative scheme are to avoid completely the interference between the primary and the secondary transmissions and to overcome power constraints when using the relay node.

The paper is organized as follows. Section II presents the system
model. Section III details the proposed cooperation scheme,  the
outage probability analysis and the power allocation problem. Section IV presents and
discusses the numerical evaluation and simulation results and finally, a
conclusion is provided in section V.

\section{System Model}
We assume a cognitive radio network where one secondary transmission can coexist with a primary transmission at the same time and on the same frequency band. A relay node, that belongs either to the primary or the secondary network, can assist both the primary and secondary transmissions. We denote by PT, PD, ST, SD and R the primary transmitter, primary destination, secondary transmitter, secondary destination and the relay node respectively. Without loss of generality we assume that R is a decode-and-forward relay node \cite{Laneman1}.

During one time slot, PT and ST send different symbols $x_p$ and $x_s$ respectively, with energy $E\{|x_{p}|^2\}=E\{|x_{s}|^2\}=1$, using transmit powers $P_p$ and $P_s$ in order to achieve data rates $R_p$ and $R_s$ respectively. We assume that the channels between different nodes are Rayleigh fading channels and are stationary during a time slot.

\section{Proposed Cooperative Scheme}

\subsection{Description}
The proposed scheme is illustrated in
Fig. \ref{Fig2_Network_R_assisted_PU}. Each time slot is divided
into two sub-slots. In the first sub-slot, R receives and attempts to decode the signals transmitted by PT and ST. Practically, a Successive Interference Cancelation (SIC) receiver could be used at R to detect $x_p$ and $x_s$ \cite{Tse}. Hence, decoding proceeds over two stages. In the first one, R decodes $x_p$ while treating $x_s$ as an interfering signal. The achieved primary Signal-plus-Interference-to-Noise-Ratio (SINR) at R is $\frac{\gamma_p |h_{pr}|^2}{\gamma_s |h_{sr}|^2+1}$, where $p$, $s$ and $r$ stand for primary users, secondary users and relay respectively, $\gamma_a=P_a/N_0$ ($a=p,s$ or $r$) is the Signal-to-Noise Ratio (SNR), $N_0$ is the variance of the zero mean Additive White Gaussian Noise (AWGN) received at $b$ and denoted $n_b$ ($b=p,s$ or $r$), and $h_{ab}$ is the fading coefficient of the channel from $a$ to $b$ with variance $\sigma_{ab}^{2}$.
In the second stage, R can subtract $x_p$ from the aggregate received signal and then decodes $x_s$. The secondary SINR at R is hence $\gamma_s |h_{sr}|^2$.
A similar mechanism can be used if R starts by decoding $x_s$.

The received signals at PD, R and SD during the first sub-slot are respectively given by:
\begin{eqnarray}
\label{eq:received_PD_direct}
y_{p}(1)&=&\sqrt{P_{p}}h_{pp}x_p+\sqrt{P_{s}}h_{sp}x_s+n_{p},\\
\label{eq:received_PT_R}
y_{r}(1)&=&\sqrt{P_{p}}h_{pr}x_p+\sqrt{P_{s}}h_{sr}x_s+n_{r},\\
\label{eq:received_SD_direct}
y_{s}(1)&=&\sqrt{P_{s}}h_{ss}x_s+\sqrt{P_{p}}h_{ps}x_p+n_{s}.
\end{eqnarray}
We assume that the secondary transmit power $P_s$ is calculated to satisfy a certain primary outage probability threshold denoted $\varepsilon$. Hence, $P_s$ can be written as (\cite{Zou},eq.(5)):
\begin{equation}
\label{eq:condition_gamma_ST}
P_s=\frac{P_p \sigma_{pp}^{2}}{\Theta_p \sigma_{sp}^{2}} \rho^{+},
\end{equation}
where $\rho^{+}=max\left(0,\rho=\frac{e^{-\frac{\Theta_p}{{\gamma}_{p}\sigma_{pp}^{2}}}}{1-\varepsilon}-1\right)$ and $\Theta_p=2^{R_p}-1$.

The participation of R on the communication is indicated by a parameter $D$ calculated as indicated below \cite{Tse}:
\begin{eqnarray}
\label{eq:condition_CR_helps_PU_SU}
\nonumber &\mathrm{If}\;\left\{A_p \cap B_p \cap C_p\right\} \cup \left\{A_s \cap B_s \cap C_s \right\},\;\mathrm{then}&\;D = 1,\\
&\mathrm{otherwise}\;&D = 0,
\end{eqnarray}
where \begin{eqnarray}
\nonumber A_i&=&\left\{\frac{1}{2}{{\log }_2}\left( {1 + \frac{{\gamma _{i}}{{\left| {{h_{ir}}} \right|}^2}}{{\gamma _{\bar{i}}}{{\left| {{h_{\bar{i}r}}} \right|}^2}+1}} \right){\rm{ \geq }}{R_i}\right\},\\
\nonumber B_i&=& \left\{ \frac{1}{2} log_2 \left( 1 + \gamma_{\bar{i}} {\left| {{h_{\bar{i} r}}} \right|}^2 \right) \geq R_{\bar{i}} \right\}\\
\mathrm{and}\;\;\;\;\;\;\;\;\;\;\;\;\;\;\; \nonumber C_i&=&\left\{   {\gamma _{i}}{{\left| {{h_{ir}}} \right|}^2} > {\gamma _{\bar{i}}}{{\left| {{h_{\bar{i}r}}} \right|}^2} \right\},
\end{eqnarray}
where $i=p$ or $s$ and $\bar{i}=s$ if $i=p$ and $\bar{i}=p$ if $i=s$.

If $D=1$, PT and ST remain silent and R
allocates a
fraction of its power $\alpha P_{r}$ ($0 \leq
\alpha \leq 1$) to send the regenerated primary signal
and the rest of its power, i.e. $(1-\alpha)P_{r}$, to send the regenerated secondary signal in the second sub-slot.

If $D=0$, then R remains silent whereas PT and ST
retransmit the same signal in the second sub-slot.
Finally, PD (and SD) combines the two received copies of the primary (secondary) signal using
Maximum Ratio Combining (MRC) and estimates the original signal using Maximum Likelihood Detection (MLD).

In the second sub-slot, two possible cases may occur. When $D=0$, the signal received at SD and its corresponding SINR are given respectively by:
\begin{eqnarray}
\label{eq:received_Rs_help_both_D0}
y_{s}(2|D=0)&=&\sqrt{P_{s}}h_{ss}x_s+\sqrt{P_{p}}h_{ps}x_p+n_{s},\\
\label{eq:SINR_SD_D0}
SINR_{s}(D=0)&=&\frac{2\gamma_{s}|h_{ss}|^2}{\gamma_{p}|h_{ps}|^2+1}.
\end{eqnarray}
When $D=1$, R assists both PT and ST. The
received signals at PD and SD are then expressed respectively by:
\begin{equation}
\label{eq:received_R_helpPU_D1}
y_{p}(2|D=1)=\sqrt{\alpha P_{r}}h_{rp}x_p+\sqrt{(1-\alpha)P_{r}}h_{rp}x_s+n_{p},
\end{equation}
\begin{equation}
\label{eq:received_R_helpPU_D1_at_SD}
y_{s}(2|D=1)=\sqrt{(1-\alpha)P_{r}}h_{rs}x_s+\sqrt{\alpha P_r}h_{rs}x_p+n_{s}.
\end{equation}

Using
(\ref{eq:received_PD_direct}) and (\ref{eq:received_R_helpPU_D1})
with MRC and (\ref{eq:received_SD_direct}) and (\ref{eq:received_R_helpPU_D1_at_SD})
with MRC, the SINR at PD and SD can be respectively written as:
\begin{equation}
\label{eq:SINR_PD_D1}
SINR_{p}(D=1)=\frac{\gamma_{p}|h_{pp}|^2}{\gamma_{s}|h_{sp}|^2+1}+\frac{\alpha \gamma_{r}|h_{rp}|^2}{(1-\alpha)\gamma_{r}|h_{rp}|^2+1},
\end{equation}
\begin{equation}
\label{eq:SINR_SD_D1_CR_help_PU}
SINR_{s}(D=1)=\frac{\gamma_{s}|h_{ss}|^2}{\gamma_{p}|h_{ps}|^2+1}+\frac{(1-\alpha) \gamma_{r}|h_{rs}|^2}{\alpha\gamma_{r}|h_{rs}|^2+1}.
\end{equation}

\subsection{Outage Probability Analysis}
In this section, we derive exact closed-form and upper bound expressions for the conditional primary and secondary outage probabilities for the proposed cooperative scheme.We shall use the following Lemmas.
\begin{Lemma}
The probability of $D=1$ is
given by:
\label{Lemma1}
\begin{equation}
\label{eq:prob_CR_help_both_D1}
P(D=1)= \frac{\tilde{\gamma}_{pr}}{\tilde{\gamma}_{pr}+\tilde{\gamma}_{sr}}e^{-\frac{M_p}{\tilde{\gamma}_{pr}}-\frac{\Lambda_s}{\tilde{\gamma}_{sr}}}+\frac{\tilde{\gamma}_{sr}}{\tilde{\gamma}_{sr}+\tilde{\gamma}_{pr}}e^{-\frac{M_s}{\tilde{\gamma}_{sr}}-\frac{\Lambda_p}{\tilde{\gamma}_{pr}}},
\end{equation}
where $\tilde{\gamma}_{ab}=\gamma_a \sigma_{ab}^{2}$ ($a=p,s$ or $r$ and $b=p,s$ or $r$), $\Lambda_p=2^{2R_p}-1$, $\Lambda_s=2^{2R_s}-1$ and $M_i=max\left( \Lambda_i(1+\Lambda_{\bar{i}}), \Lambda_{\bar{i}} \right)$ with $i=p$ or $s$ and $\bar{i}=s$ if $i=p$ and $\bar{i}=p$ if $i=s$.
\end{Lemma}
\begin{IEEEproof}
Since $\left\{ A_p \cap B_p \cap C_p \right\}$ and $\left\{ A_s \cap B_s \cap C_s \right\}$ are independent, then
\begin{eqnarray}
\label{eq:PD1}
\nonumber P(D=1)&=&\sum_{i=\left\{ p,s\right\} } P(A_i\cap B_i \cap C_i)\\
&=&\sum_{i=\left\{ p,s\right\} } P(A_i\cap B_i | C_i)P(C_i).
\end{eqnarray}
The random variable $|h_{ab}|^2$ has an exponential distribution with parameter $1/\sigma_{ab}^2$, thus we get:
\begin{equation}
\label{eq:PC}
P(C_i)=\frac{\tilde{\gamma}_{ir}}{\tilde{\gamma}_{ir}+\tilde{\gamma}_{\bar{i}r}},
\end{equation}
and since $|h_{ir}|^2$ and $|h_{\bar{i}r}|^2$ are independent
\begin{eqnarray}
\label{eq:PABC}
\nonumber P(A_i\cap B_i | C_i)&=&P(\gamma_i|h_{ir}|^2 \geq \Lambda_i(1+\Lambda_{\bar{i}})\\
\nonumber &{}&\mathrm{and}\; \gamma_i|h_{ir}|^2 \geq \Lambda_{\bar{i}} \;\mathrm{and}\; \gamma_{\bar{i}}|h_{\bar{i}r}|^2 \geq \Lambda_{\bar{i}} )\\
\nonumber &=&P(\gamma_i|h_{ir}|^2 \geq M_i)P(\gamma_{\bar{i}}|h_{\bar{i}r}|^2 \geq \Lambda_{\bar{i}})\;\\
&=&e^{-\frac{M_i}{\tilde{\gamma}_{ir}}}e^{-\frac{\Lambda_{\bar{i}}}{\tilde{\gamma}_{\bar{i}r}}}.
\end{eqnarray}
By combining (\ref{eq:PC}) and (\ref{eq:PABC}) in (\ref{eq:PD1}), we obtain (\ref{eq:prob_CR_help_both_D1}). This completes the proof of Lemma.\ref{Lemma1}.
\end{IEEEproof}
The probability of
retransmitting $x_p$ and $x_s$ by PT and ST respectively in the second sub-slot is $P(D=0)=1-P(D=1)$. The received SINR at SD is given by (\ref{eq:SINR_SD_D0}) and the conditional secondary outage probability can be given by (\cite{Zou},eq.(21)):
\begin{equation}
P_{sec}(outage|D=0)=P(SINR_s<\Lambda_s)=1-\frac{2
\tilde{\gamma}_{ss}e^{-\frac{\Lambda_s}{2 \tilde{\gamma}_{ss}}}}{2
\tilde{\gamma}_{ss}+\Lambda_s \tilde{\gamma}_{ps}}.
\end{equation}

If $D=1$, only the relay node transmits in the second sub-slot. The received SINR at PD and SD are given by (\ref{eq:SINR_PD_D1}) and (\ref{eq:SINR_SD_D1_CR_help_PU}) respectively. The outage probabilities depend on $\alpha$, the fraction of $P_r$ to be allocated to transmit $x_p$, and can be calculated using the following Lemmas \ref{Lemma2} and \ref{Lemma3}.
\begin{Lemma}
\label{Lemma2}
If $\alpha=0 \;\mathrm{or}\;\alpha=1$, R allocates all its power to transmit only one signal. Then, the conditional primary and secondary outage probabilities are given by:
\begin{eqnarray}
\label{eq:outage_prob_pri_CR_help_both_D1}
&P_{pri}(outage|D=1)=&\\
&\nonumber \left\{ {\begin{array}{*{20}{c}}
  {1-\frac{
\tilde{\gamma}_{pp}e^{-\frac{\Lambda_p}{\tilde{\gamma}_{pp}}}}{
\tilde{\gamma}_{pp}+\Lambda_p \tilde{\gamma}_{sp}},}&{\mathrm{if}\;\alpha=0} \\
  {1-e^{-\frac{\Lambda_p}{\tilde{\gamma}_{rp}}}\left(
1+\frac{\tilde{\gamma}_{pp}}{\tilde{\gamma}_{sp}\tilde{\gamma}_{rp}}e^{-\frac{\tilde{\gamma}_{pp}}{\tilde{\gamma}_{sp}}\left(
\frac{1}{\tilde{\gamma}_{rp}}-\frac{1}{\tilde{\gamma}_{pp}}
\right)} \Gamma_p \right),}&{\mathrm{if}\;\alpha=1}
\end{array}} \right.&
\end{eqnarray}
\begin{eqnarray}
\label{eq:outage_prob_sec_CR_help_both_D1} &P_{sec}(outage|D=1)=&\\
&\nonumber \left\{ {\begin{array}{*{20}{c}}
{1-e^{-\frac{\Lambda_s}{\tilde{\gamma}_{rs}}}\left(
1+\frac{\tilde{\gamma}_{ss}}{\tilde{\gamma}_{ps}\tilde{\gamma}_{rs}}e^{-\frac{\tilde{\gamma}_{ss}}{\tilde{\gamma}_{ps}}\left(
\frac{1}{\tilde{\gamma}_{rs}}-\frac{1}{\tilde{\gamma}_{ss}}
\right)} \Gamma_s \right),}&{\mathrm{if}\;\alpha=0}\\
{1-\frac{
\tilde{\gamma}_{ss}e^{-\frac{\Lambda_s}{\tilde{\gamma}_{ss}}}}{
\tilde{\gamma}_{ss}+\Lambda_s \tilde{\gamma}_{ps}},}&{\mathrm{if}\;\alpha=1}
\end{array}} \right.&
\end{eqnarray}
where \[ \Gamma_p=\int_{{{\tilde \gamma }_{pp}}}^{{{\tilde
\gamma }_{pp}} + {\Lambda _p}{{\tilde \gamma }_{sp}}}
{\frac{e^{\frac{x}{{{{\tilde \gamma }_{sp}}}}\left(
{\frac{1}{{{{\tilde \gamma }_{rp}}}} - \frac{1}{{{{\tilde
\gamma }_{pp}}}}} \right)}}{x}dx},\] and \[\Gamma_s=\int_{{{\tilde \gamma }_{ss}}}^{{{\tilde
\gamma }_{ss}} + {\Lambda _s}{{\tilde \gamma }_{ps}}}
{\frac{e^{\frac{x}{{{{\tilde \gamma }_{ps}}}}\left(
{\frac{1}{{{{\tilde \gamma }_{rs}}}} - \frac{1}{{{{\tilde
\gamma }_{ss}}}}} \right)}}{x}dx}\] can be calculated using the mathematical tables in \cite{Beyer}.
\end{Lemma}
\begin{IEEEproof}
If $\alpha=0$, $SINR_p(D=1)=\frac{\gamma_p|h_{pp}|^2}{\gamma_s|h_{sp}|^2+1}$\\ and $P_{pri}(outage|D=1)$ is given by (\ref{eq:outage_prob_pri_CR_help_both_D1}). \\
If $\alpha=1$, $SINR_p(D=1)=\frac{\gamma_p|h_{pp}|^2}{\gamma_s|h_{sp}|^2+1}+\gamma_r|h_{rp}|^2$\\ and $P_{pri}(outage|D=1)$ is expressed by:
\begin{eqnarray}
\label{eq:p_out_sec_CR_help_both_demo}
P_{pri}(outage|D=1)&=&P(v<\Lambda_p - w)\\
\nonumber &=&\int _{0}^{\Lambda_p}{f_w(w)} \int_{0}^{\Lambda_p
 - w} f_v(v) dv dw,
\end{eqnarray}
where
$v=\frac{\gamma_{p}|h_{pp}|^2}{\gamma_{s}|h_{sp}|^2+1}$
and $\omega=\gamma_{r}|h_{rp}|^2$. We have:
\begin{equation}
\label{eq:integ}
\int_{0}^{\Lambda_p - w} f_v(v)
dv=1-\frac{\tilde{\gamma}_{pp}e^{-\frac{\Lambda_p-w}{\tilde{\gamma}_{pp}}}}{\tilde{\gamma}_{pp}+\left(\Lambda_p-w\right)\tilde{\gamma}_{sp}}.
\end{equation}
Then,
\begin{eqnarray}
 P_{pri}(outage|D=1)=1-e^{-\frac{\Lambda_p}{\tilde{\gamma}_{rp}}}\\
\nonumber
-\frac{\tilde{\gamma}_{pp}e^{-\frac{\Lambda_p}{\tilde{\gamma}_{rp}}}}{\tilde{\gamma}_{sp}\tilde{\gamma}_{rp}}
\int_ {\tilde{\gamma}_{pp}}
 ^{\tilde{\gamma}_{pp}+\Lambda_s \tilde{\gamma}_{sp}}
\frac{1}{\varphi} e^{\frac{\varphi
-{\tilde{\gamma}_{pp}}}{\tilde{\gamma}_{sp}} \left(
\frac{1}{\tilde{\gamma}_{rp}}-\frac{1}{\tilde{\gamma}_{pp}}
\right)}d\varphi,
\end{eqnarray}
where we performed a variable change
$\varphi=\tilde{\gamma}_{pp}+\left( \Lambda_p - w
\right)\tilde{\gamma}_{sp}$. Hence, $P_{pri}(outage|D=1)$ is
given by (\ref{eq:outage_prob_pri_CR_help_both_D1}). Similar calculation can be made to prove (\ref{eq:outage_prob_sec_CR_help_both_D1}). This completes the proof of Lemma.\ref{Lemma2}.
\end{IEEEproof}
Finally,the secondary outage probability $P_{out_{sec}}$ can be written as (\cite{Zou},eq.(28)):
\begin{eqnarray}
\label{eq:p_out_sec}
\nonumber P_{out_{sec}}&=&P(D=0)P_{sec}(outage|D=0)\\
&+&P(D=1)P_{sec}(outage|D=1).
\end{eqnarray}
\begin{Lemma}
\label{Lemma3} If $0 < \alpha < 1$, then the conditional
primary and secondary outage probabilities are upper bounded by $U_p$ and $U_s$ respectively, given by:
\begin{eqnarray}
\label{eq:UP_pri}
U_p=&\left\{ {\begin{array}{*{20}{c}}
{ {1 - \frac{{{{\tilde \gamma }_{pp}}{e^{ - \frac{{{\Lambda _p}}}{{{{\tilde \gamma }_{pp}}}}}}}}{{{{\tilde \gamma }_{pp}} + {\Lambda _p}{{\tilde \gamma }_{sp}}}}},}&{{\alpha} \le \frac{\Lambda_p }{{1 + \Lambda_p }}}\\
{\left( {1 - \frac{{{{\tilde \gamma }_{pp}}{e^{ -
\frac{{{\Lambda _p}}}{{{{\tilde \gamma }_{pp}}}}}}}}{{{{\tilde \gamma }_{pp}} + {\Lambda _p}{{\tilde
\gamma }_{sp}}}}} \right)\left( {1 - {e^{ - \frac{{{\Lambda
_p}}}{{{{\tilde \gamma }_{rp}}(\alpha  - (1 - \alpha ){\Lambda
_p})}}}}} \right),}&{{\alpha} \ge \frac{\Lambda_p }{{1 + \Lambda_p
}}}
\end{array}} \right.&
\end{eqnarray}
\begin{eqnarray}
\label{eq:UP_sec}
U_s=&\left\{ {\begin{array}{*{20}{c}}
{1 - \frac{{{{\tilde \gamma }_{ss}}{e^{ - \frac{{{\Lambda _s}}}{{{{\tilde \gamma }_{ss}}}}}}}}{{{{\tilde \gamma }_{ss}} + {\Lambda _s}{{\tilde \gamma }_{ps}}}},}&{{\alpha} \ge \frac{1}{1+\Lambda_s }}\\
{\left( {1 - \frac{{{{\tilde \gamma }_{ss}}{e^{ -
\frac{{{\Lambda _s}}}{{{{\tilde \gamma }_{ss}}}}}}}}{{{{\tilde \gamma }_{ss}} + {\Lambda _s}{{\tilde
\gamma }_{ps}}}}} \right)\left( {1 - {e^{ - \frac{{{\Lambda
_s}}}{{{{\tilde \gamma }_{rs}}\left( {1 - \alpha  - \alpha
{\Lambda _s}} \right)}}}}} \right),}&{{\alpha} \le \frac{{1 }}{1 +
\Lambda_s }}
\end{array}} \right.&
\end{eqnarray}
\end{Lemma}
\begin{IEEEproof}
Using (\ref{eq:SINR_PD_D1}), the conditional primary
outage probability when $0<\alpha<1$ is
given by:
\begin{equation}
\label{eq:outage_demo}
P_{pri}(outage|D=1)=\int_{0}^{\Lambda_p} {f_z(z)} \int
_{0}^{\Lambda_p-z} {f_q(q)}dq dz,
\end{equation}
where
$q=\frac{\gamma_{p}|h_{pp}|^2}{\gamma_{s}|h_{sp}|^2+1}$
and $z=\frac{\alpha \gamma_{r}|h_{rp}|^2}{(1-\alpha)
\gamma_{r}|h_{rp}|^2+1}$.  Similarly to
(\ref{eq:integ}), we obtain:
\begin{equation}
\label{f_qq} \int _{0}^{\Lambda_p - z}
{f_q(q)}dq=1-\frac{\tilde{\gamma}_{pp}e^{-\frac{\Lambda_p
-z}{\tilde{\gamma}_{pp}}}}{\tilde{\gamma}_{pp}+\left(
\Lambda_p -z \right) \tilde{\gamma}_{sp}}, \forall z \leq
\Lambda_p
\end{equation}
where $z$ is a random variable that is a function of\\ $w=\gamma_{r}
|h_{rp}|^2$. Let $g$ be the function given by:
\begin{equation}
z=g(w)=\frac{\alpha w}{(1-\alpha)w+1}.
\end{equation}
The probability density function of $z$ is given by:
\begin{eqnarray}
\label{f_rr} \nonumber
f_z(z)&=&\frac{1}{|g'(g^{-1}(z))|}f_w(g^{-1}(z))\\
 &=&\frac{\alpha
e^{-\frac{z}{\left( \alpha-\left( 1-\alpha\right)z
\right)\tilde{\gamma}_{rp}}}}{\left( \alpha-\left(
1-\alpha\right)z \right)^2 \tilde{\gamma}_{rp}},\forall z \leq
\frac{\alpha}{1-\alpha}.
\end{eqnarray}
Substituting (\ref{f_qq}) and (\ref{f_rr}) in (\ref{eq:outage_demo}), we obtain:
\begin{eqnarray}
\label{eq:outage_demo_int}
P_{pri}(outage|D=1)=\int _{0}
^{min(\Lambda_p, \frac{\alpha}{1-\alpha} )} {f_z(z)dz}\\ \nonumber
- \int _{0}
^{min(\Lambda_p, \frac{\alpha}{1-\alpha} )} f_z(z)
\frac{\tilde{\gamma}_{pp}e^{-\frac{\Lambda_p
-z}{\tilde{\gamma}_{pp}}}}{\tilde{\gamma}_{pp}+\left(
\Lambda_p -z \right) \tilde{\gamma}_{sp}} dz.
\end{eqnarray}
Let $m=min(\Lambda_p, \frac{\alpha}{1-\alpha})$. Since $0 \leq z
\leq m$, then:
\begin{equation}
\label{borne_sup} \psi=-\frac{\tilde{\gamma}_{pp}e^{-\frac{\Lambda_p
-z}{\tilde{\gamma}_{pp}}}}{\tilde{\gamma}_{pp}+\left(
\Lambda_p -z \right) \tilde{\gamma}_{sp}}  \leq
-\frac{\tilde{\gamma}_{pp}e^{-\frac{\Lambda_p}{\tilde{\gamma}_{pp}}}}{\tilde{\gamma}_{pp}+
\Lambda_p \tilde{\gamma}_{sp}}.
\end{equation}
Combining the expression of $m$ and the upper bound of $\psi$ into (\ref{eq:outage_demo_int}), we
obtain the upper bound $U_p$ given by (\ref{eq:UP_pri}). A similar proof can be used for (\ref{eq:UP_sec}). This completes the proof of Lemma.\ref{Lemma3}.
\end{IEEEproof}
Consequently, $P_{out_{sec}}$ is upper bounded by $U'_s$ that can be written as:
\begin{eqnarray}
\label{eq:UP_sec_out} U'_s=P(D=0)P_{sec}(outage|D=0)+P(D=1)U_s.
\end{eqnarray}

\subsection{Power Allocation}
\begin{table}
\caption{$\alpha_{\varepsilon}$ and $U'_s$ versus $\varepsilon$}
\begin{center}
\label{table:alpha_fct_threshold}
\begin{tabular}
{| l || c | c |c |c |c |c |c |c |c |c |r | } \hline
$\varepsilon$ &  0.04& 0.05 & 0.06& 0.07 & 0.08& 0.09 \\ \hline
$\alpha_{\varepsilon}$ & 0,488 & 0,489 & 0,489 & 0,488&	0,488&	0,487 \\ \hline
$U'_s$ & 0,021&	0,016&	0,012&	0,01&	0,009&	0,007 \\ \hline
\end{tabular}
\end{center}
\end{table}
We formulate the power allocation problem as follows;
\begin{equation}
\label{eq:formulated_prob_CR_help_both}
\mathop {\min }\limits_{\alpha ,{\gamma _{r}}} U'_s\;
\mathrm{s.t}\; U_p \le \varepsilon, \; \forall \; 0<\alpha<1.
\end{equation}
We make use of a simple suboptimal solution that consists of extracting $\alpha$ ($0<\alpha<1$) as a function of $\gamma_{r}$ and $U_p$ from (\ref{eq:UP_pri}). Hence, $\forall \; \alpha \geq \frac{\Lambda_p}{1+\Lambda_p}$
\begin{equation}
\label{eq:alpha_fct_upper_bound}
\alpha=\frac{\Lambda_p}{1+\Lambda_p} \left[ { 1-{\frac{1}{\tilde{\gamma}_{rp}} ln\left( 1-\frac{U_p}{1-\frac{\tilde{\gamma}_{pp}e^{-\frac{\Lambda_p}{\tilde{\gamma}_{pp}}}}{\tilde{\gamma}_{pp}+\Lambda_p \tilde{\gamma}_{sp}}} \right)^{-1}} } \right].
\end{equation}
The case of $\alpha < \frac{\Lambda_p}{1+\Lambda_p}$ is not considered since $U_p$ does not depend on $\alpha$ and $\gamma_r$. When $U_p=\varepsilon$ in (\ref{eq:alpha_fct_upper_bound}), the associated value of $\alpha$, $\alpha_{\varepsilon}$ and $U'_s({\alpha_{\varepsilon}})$ can be obtained as shown in Table \ref{table:alpha_fct_threshold}. We choose arbitrarily $R_p=0.4bits/s/Hz$, $R_s=0.2bits/s/Hz$, $\gamma_{p}=20dB$ and $\gamma_{r}=10dB$. We assume also that
$\sigma_{pp}^2=\sigma_{ss}^2=\sigma_{pr}^2=\sigma_{rp}^2=\sigma_{sr}^2=\sigma_{sr}^2=1$, and $\sigma_{ps}^2=\sigma_{sp}^2=0.1$. As $\varepsilon$ is less severe, more power is allocated to send $x_s$ and thus $U'_s$ is improved.
\begin{figure}
  \centering
  \includegraphics[width=260pt]{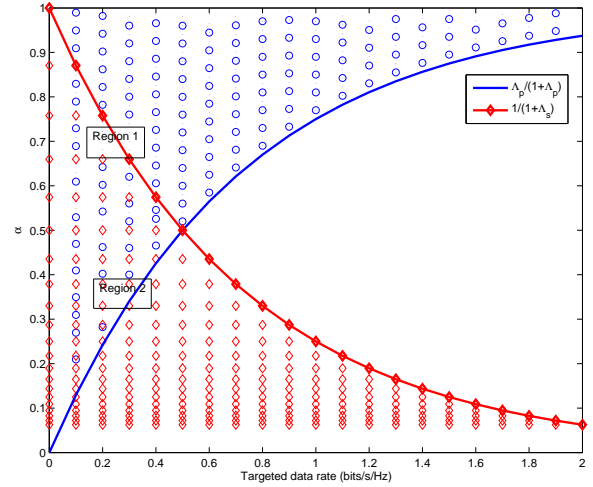}
  \caption{$R_p$ and $R_s$ versus $\alpha$}
  \label{Fig_behavior_upper_bounds_alpha2}
\end{figure}

Furthermore, $U_{p}$ and $U_{s}$ depend on $R_p$, $R_s$
and $\alpha$. In Fig. \ref{Fig_behavior_upper_bounds_alpha2}, we
illustrate the relationship between these parameters. Regions 1 (circles) and 2 (diamonds) correspond to $U_p(\alpha,\gamma_{r})$ and $U_s(\alpha,\gamma_{r})$ respectively. In other words, adequate values of $\alpha$ and
$\gamma_{r}$ could improve $P_{out_{pri}}$ and/or
$P_{out_{sec}}$ in these regions. But, since $U_p$
should respect $\varepsilon$ and since we aim to improve $U'_s$, the common region for regions 1 and 2 is the best choice (circles+diamonds). Hence, our
proposed scheme performs best when $R_p<0.5
bits/s/Hz$ and $R_s<0.5
bits/s/Hz$. Fig. \ref{Fig_behavior_upper_bounds_alpha2} can also serve as
a reference map to optimally choose $\alpha$ or $R_s$ at fixed $R_p$. For instance, $(R_p,\alpha)=(1,0.76)$ corresponds to $R_s=0.2 bits/s/Hz$ and $(R_p,R_s)=(0.4,0.2)$ to $\alpha \in [0.43,0.75]$.

\section{Numerical and Simulation Results}
In this section we evaluate $P_{out_{sec}}$ using the proposed cooperation scheme based on (\ref{eq:p_out_sec}) and (\ref{eq:UP_sec_out}). We also compare it to the non-cooperative scheme and to the relay-assisted secondary user scheme proposed in \cite{Zou}. We assume that $\varepsilon=0.03$, $R_p=0.4 bits/s/Hz$ and $R_s=0.2 bits/s/Hz$ and we denote by $\mu_1=\sigma_{pr}^2=\sigma_{rp}^2$ and $\mu_2=\sigma_{sr}^2=\sigma_{rs}^2$.

\begin{figure}
  \includegraphics[width=260pt]{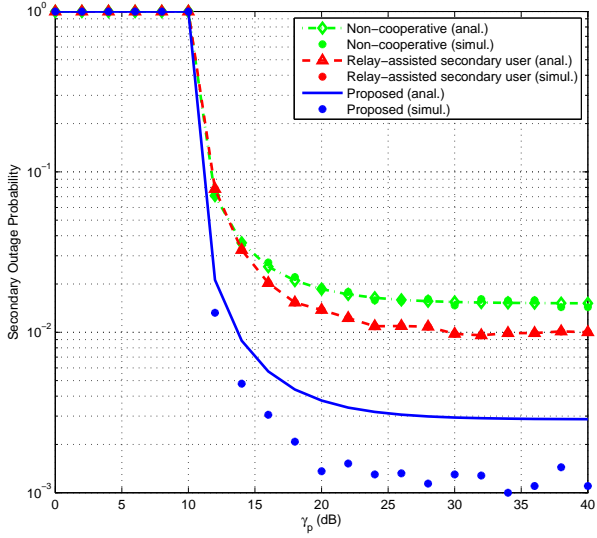}
  \caption{Secondary outage probability $P_{out_{sec}}$ versus $\gamma_{p}$ with
$\sigma_{pp}^2=\sigma_{ss}^2=1$,
$\sigma_{ps}^2=\sigma_{sp}^2=0.1$, $\mu_1=\mu_2=1$ and $\alpha=0.5$}
  \label{Fig_all_techniques_comparison_3}
\end{figure}
In Fig. \ref{Fig_all_techniques_comparison_3} we assume $\mu_1=\mu_2=1$. The relay-assisted secondary user scheme and the proposed scheme
provide better outage probability performances than the
non-cooperative one, with a preference for the proposed scheme. Since $\mu_1=\mu_2=1$, it is more likely that R will transmit in the second sub-slot and thus improves on the average the outage probability of the secondary system with respect to $\varepsilon$. However, when R helps ST only, it causes an important interference on PD ($\mu_1=1$). Moreover, all the schemes present a cut-off point at $\gamma_p=12dB$ below which no secondary transmissions are allowed \cite{Zou}.

We see that the numerical results match the simulation ones for the non-cooperative and the relay-assisted secondary user scheme and upper bound the ones for the proposed scheme. This gap between the numerical and simulation results for the proposed scheme is explained by two factors. First, we provide an upper bound for $P_{out_{sec}}$. Second, to get $\alpha$ and $\gamma_r$ in the power allocation problem, we use an upper bound for $P_{out_{pri}}$.

\begin{figure}
  \includegraphics[width=260pt]{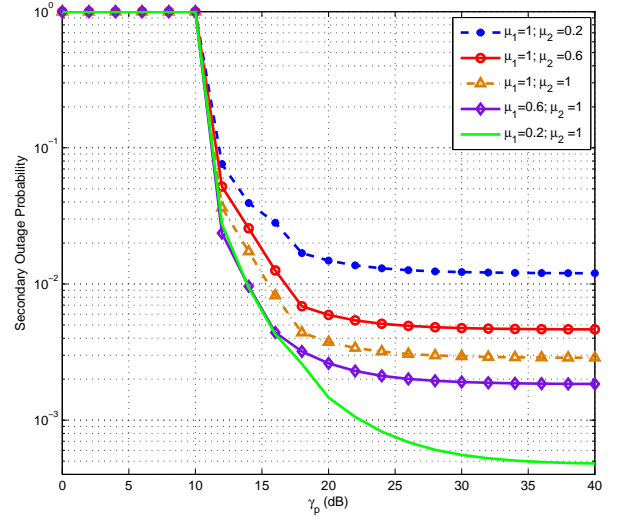}
  \caption{Secondary outage probability $P_{out_{sec}}$ versus $\gamma_{p}$ with
$\sigma_{pp}^2=\sigma_{ss}^2=1$,
$\sigma_{ps}^2=\sigma_{sp}^2=0.1$ and $\alpha=0.5$}
  \label{Fig_CR_helps_both_different_channel_condition}
\end{figure}

\begin{figure}
  \includegraphics[width=260pt]{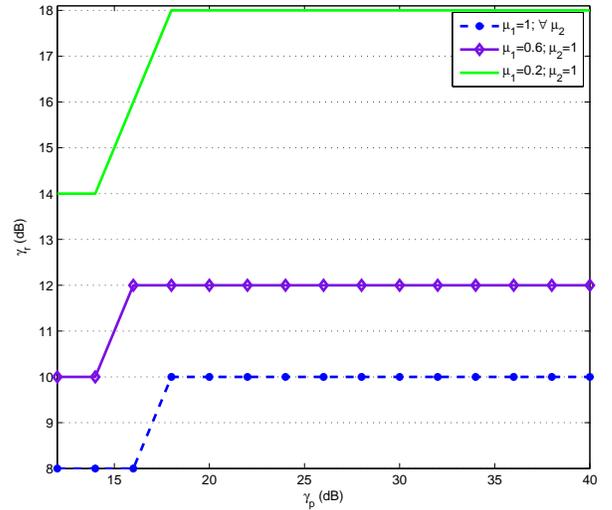}
  \caption{$\gamma_r$ versus $\gamma_p$ for different channel conditions}
  \label{subfig1}
\end{figure}

In Fig. \ref{Fig_CR_helps_both_different_channel_condition} we illustrate the numerical results of
$P_{out_{sec}}$ versus $\gamma_{p}$ for the proposed scheme at fixed $\alpha$ under different channel conditions.
We see that when $\mu_2=1$ and $\mu_1$ decreases, $P_{out_{sec}}$ drops off. However, at fixed $\mu_1=1$ and decreasing $\mu_2$, $P_{out_{sec}}$ degrades. We see that the condition of ST-R and R-SD channels has a more important impact on $P_{out_{sec}}$ than PT-R and R-PD channels condition.

In Fig. \ref{subfig1}, we plot the power consumed by R for constant $\alpha$. When $\mu_1=1$, the same low power is used by R for any $\mu_2$ value. Hence, $\varepsilon$ is respected and no more power is needed. However, as $\mu_1$ decreases, $\gamma_r$ increases drastically. This augmentation of $\gamma_r$ compensates the degradation of PT-R and R-PD channels in order to respect $\varepsilon$. Even though this is an important cost on $\gamma_r$, it provides a significant improvement in the outage performance of the secondary transmission.

\begin{figure}
  \includegraphics[width=260pt]{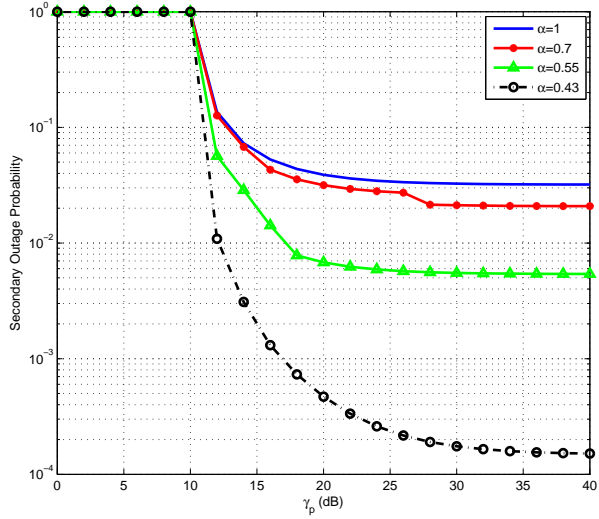}
  \caption{Secondary outage probability $P_{out_{sec}}$ versus $\gamma_{p}$ with
$\sigma_{pp}^2=\sigma_{ss}^2=1$,
$\sigma_{ps}^2=\sigma_{sp}^2=0.1$ and $\mu_1=\mu_2=1$}
  \label{Fig_CR_helps_both_different_alpha_error}
\end{figure}

Fig. \ref{Fig_CR_helps_both_different_alpha_error} shows the numerical results of the
impact of $\alpha$ on $P_{out_{sec}}$.
For each ($\alpha$, $\gamma_{p}$) we search for the minimum value of
$\gamma_{r}$ to reach $\varepsilon$. As $\alpha$ decreases
from 1 to 0.43, $P_{out_{sec}}$ improves significantly. Since
$\mu_1=\mu_2=1$, allocating more power to
assist the secondary users is beneficial. When $\alpha=1$, all the power is assigned to
assist the primary users which causes the secondary outage
probability to increase rapidly (ST is silent in the second sub-slot). The case of $\alpha<\frac{\Lambda_p}{1+\lambda_p}=0.43$ is not
plotted since in this scenario $\varepsilon$ is not respected and $P_{out_{sec}}=1$ for any $\gamma_p$.
The results showed also that $\gamma_r$ increases rapidly as $\alpha$ drops off. The power increase balances the loss on the power allocated for the primary signal.

\section{Conclusion}
In this paper, we proposed a novel cooperative transmission scheme in underlay cognitive radio networks, where a relay node is able to assist simultaneously the primary and secondary users. We derived exact closed-form and upper bound expressions for the conditional primary and secondary outage probabilities in a Rayleigh fading environment. Moreover, we have investigated the power allocation problem and proposed a simple allocation algorithm. We evaluated numerically and by simulation our proposed cooperation and we showed the obtained performance improvements.

\newpage

\bibliographystyle{IEEEtran}
\bibliography{IEEEabrv,tau}

\end{document}